\documentclass[%
 a4,
 reprint,
 preprintnumbers,
 nofootinbib,
 amsmath,amssymb,
 aps,
 prd,
]{revtex4-2}

\usepackage{amsmath}
\usepackage{amssymb}

\usepackage{graphicx}% Include figure files
\usepackage{dcolumn}% Align table columns on decimal point
\usepackage{bm}% bold math
\usepackage{hyperref}% add hypertext capabilities
\usepackage[normalem]{ulem} % for \sout{}

\usepackage{booktabs} % Make tables bearable to look at
\usepackage[svgnames,table]{xcolor}
\usepackage[caption=false]{subfig}

\usepackage{mathrsfs} % Lagrangian (\mathscr{L})

\usepackage{bm} % bold math
\usepackage{color} % for coloured text

\newcommand\MSbar{$\overline{\rm MS}$}
\newcommand\op{\theta_{\mathrm{op}}}
\newcommand\crit{\theta_{\mathrm{c}}}

\graphicspath{
  {./}
}

\begin{document}

\preprint{HIP-2024-9/TH}

% Just an idea:
\title{A nonperturbative test of nucleation calculations for strong phase transitions}

\newcommand{\HIPetc}{\affiliation{
Department of Physics and Helsinki Institute of Physics,
PL 64, % (Gustaf H\"{a}llstr\"{o}min katu 2),
FI-00014 University of Helsinki,
Finland
}}

\newcommand{\Nottingham}{\affiliation{
    School of Physics and Astronomy,
    University of Nottingham,
    Nottingham NG7 2RD,
    U.K.
}}

\author{Oliver Gould}
\email{oliver.gould@nottingham.ac.uk}
\Nottingham

\author{Anna Kormu}
\email{anna.kormu@helsinki.fi}
\HIPetc

\author{David J. Weir}
\email{david.weir@helsinki.fi}
\HIPetc

\date{April 2, 2024}% It is always \today, today,
             %  but any date may be explicitly specified

\begin{abstract}
Nucleation rate computations are of broad importance in particle physics and cosmology. Perturbative calculations are often used to compute the nucleation rate $\Gamma$, but these are incomplete. We perform nonperturbative lattice simulations of nucleation in a scalar field theory with a tree-level barrier, computing a final result extrapolated to the thermodynamic and continuum limits. Although the system in question should be well-described by a complete one-loop perturbative calculation, we find only qualitative agreement with the full perturbative result, with a 9\%
discrepancy in $|\log \Gamma|$.%
\footnote{
  The dynamical factor was missing in previous versions.
  The original value for the discrepancy was 20\%;
  see the erratum at the end of this paper.
}
Our result motivates further testing of the current nucleation paradigm.
\end{abstract}

\maketitle

%%%%%%%%%%%%%%%%%%%%%%%%%%%%%%% Introduction %%%%%%%%%%%%%%%%%%%%%%%%%%%%%%%
\section{Introduction}

First order phase transitions in the early universe have been of sustained interest. 
A primordial phase transition could help to explain the matter-antimatter asymmetry of the universe~\cite{Kuzmin:1985mm, Cline:2021iff, Ellis:2022lft}.
The phase transition itself, as well as any resulting nonequilibrium physics of the primordial plasma, would produce a stochastic gravitational wave background that could potentially be detected~\cite{Hindmarsh:2020hop}. 

Hints of a gravitational wave background have been seen by the various Pulsar Timing Array (PTA) collaborations~\cite{NANOGrav:2023gor, EPTA:2023fyk, Xu:2023wog, Reardon:2023gzh}. Such a gravitational wave background is expected to be due to supermassive black holes~\cite{NANOGrav:2023hfp}, but it is not possible to rule out other new physics such as a phase transition~\cite{NANOGrav:2023hvm}. Future missions such as LISA will be well placed to look for gravitational waves from electroweak-scale phase transitions~\cite{Caprini:2019egz,LISACosmologyWorkingGroup:2022jok}, but the theoretical uncertainties in the predicted gravitational wave power spectrum will need to be constrained.

Thanks to large-scale simulations~\cite{Hindmarsh:2017gnf,Jinno:2022mie}, considerable progress has been made in modelling the resulting gravitational wave power spectrum~\cite{Konstandin:2017sat,Hindmarsh:2019phv}. 
At the same time, the accuracy of the quantities which parametrise the power spectrum --  including the nucleation rate -- have faced growing scrutiny~\cite{Kainulainen:2019kyp, Gould:2019qek, Ekstedt:2020abj, Croon:2020cgk, Gould:2021oba, Ekstedt:2022ceo}.

Experimental tests of nucleation theory in condensed matter systems show a mixed picture. For the AB transition in superfluid $^3$He, there is a longstanding and puzzling discrepancy~\cite{schiffer1995nucleation, Tian:2022dzv, Hindmarsh:2024ptr}, whereas good agreement was found for nucleation in a ferromagnetic superfluid~\cite{Zenesini:2023afv}. Recently, there have been proposals to test bubble nucleation in ultracold atomic gases~\cite{Fialko:2016ggg, Billam:2018pvp,Jenkins:2023eez}.

There is a need for reliable predictions with controlled systematic uncertainties so that we can test particle physics models against gravitational wave observations, as well as for comparison with analogue experiments.
We focus on calibrating the accuracy of nucleation rate calculations in relativistic finite-temperature field theory, comparing perturbative semiclassical methods against direct nonperturbative numerical simulation.

At some critical temperature $T_c$, the effective potential energy develops two or more degenerate minima as a function of some order parameter, but the system remains in the higher energy metastable phase even after it is energetically less favourable, as the temperature continues to fall. Eventually this will lead into bubble nucleation, where bubbles of the stable phase form.

In field theory at zero temperature a metastable vacuum state can decay via quantum tunnelling~\cite{Kobzarev:1974cp}. A complete one-loop calculation of the rate of this process was first carried out by Coleman and Callan~\cite{Coleman:1977py, Callan:1977pt}, using a saddlepoint approximation of the path integral. This result has since been rederived from a number of perspectives and formalisms~\cite{Plascencia:2015pga, Andreassen:2016cff, Ai:2019fri}.

At high temperature, bubbles can nucleate classically. Within classical field theory, a complete one-loop calculation for this process was first carried out by Langer~\cite{Langer:1967ax, Langer:1969bc}, extending the tree-level theory of Cahn and Hilliard~\cite{cahn1959free}. However, the analogous calculation within quantum field theory is much less clear. 
Early papers by Linde~\cite{Linde:1980tt} and Affleck~\cite{Affleck:1980ac} gave slightly different expressions in the high temperature regime.
Neither agrees with the classical result of Langer.

The above methods all depend on a semiclassical picture of the bubble and its fluctuations. A fully numerical, lattice calculation of the bubble nucleation rate was introduced in Ref.~\cite{Moore:2000jw}. At the time, the focus was on the physics of the minimal Standard Model, where a first-order phase transition would have arisen from radiative corrections to the quartic Higgs potential.
In Ref.~\cite{Moore:2001vf} a toy model with similar features was studied. Recently a more complete study of nucleation in the minimal Standard Model was carried out ~\cite{Gould:2022ran}, motivated by the idea that any sufficiently heavy new particles could be integrated out~\cite{Gould:2019qek}. 

Another approach has been to study thermal nucleation directly as a real time process~\cite{Grigoriev:1988bd, valls1990nucleation,Alford:1993zf, Alford:1993ph, Borsanyi:2000ua, Batini:2023zpi, Pirvu:2023plk}.
This involves evolving the lattice in time, waiting until a growing bubble appears and recording the time taken for the bubble to appear. Since the nucleation rate can vary over many orders of magnitude, this approach is only viable over a narrow temperature range.

We consider what we call the real scalar theory~\cite{Gould:2021dzl}, with Lagrangian
\begin{align} \label{eq:lagrangian}
     \mathscr{L} &= -\frac{1}{2}\partial_{\mu}\varphi\partial^{\mu}\varphi-V(\varphi) - J_1 \varphi - J_2 \varphi^2, \\
     V(\varphi) &= \sigma \varphi +\frac{1}{2}m^2\varphi^2+\frac{1}{3!}g\varphi^3+\frac{1}{4!}\lambda\varphi^4,
     \label{eq:potential_4d}
\end{align}
where $\sigma$, $m^2$, $g$ and $\lambda$ are the model parameters, $J_1$ and $J_2$ represent operators of other fields, and we have used mostly plus signature. Such a scalar field may couple to the Standard Model, through Higgs portal interactions (with $J_1 = \kappa_1 H^\dagger H$ and $J_2 = \kappa_2 H^\dagger H /2$) or serve as a toy model in itself to test key ideas.

Compared with radiatively-induced transitions, the presence of the tree-level barrier in equation~\eqref{eq:potential_4d} is expected to allow for stronger transitions~\cite{Chung:2012vg}.
The perturbative expansion is also simpler~\cite{Ekstedt:2022zro, Gould:2023ovu}, allowing us to focus on bubble nucleation without extraneous details.

We focus on the dynamics of a single bubble and its real time evolution from the metastable phase to the stable phase.
The bubbles arise as long wavelength thermal fluctuations of the scalar field. 
These fluctuations are highly occupied in a thermal bath, and so their effective evolution is classical~\cite{Aarts:1997kp, Bodeker:1996wb, Greiner:1996dx}. The effects of short wavelength fluctuations appear both through screening of the effective parameters, and as noise and damping in the evolution equations~\cite{Greiner:1996dx, Morikawa:1986rp}.

Thermal fluctuations significantly modify the scalar effective potential for temperatures $T^2 \sim m^2 / \lambda$ and above. At such temperatures, modes with energies of order $m \sim \sqrt{\lambda}T$ satisfy the following Langevin equation,
\begin{align}
  \partial_t\phi(t,\mathbf{x}) &= \pi(t, \mathbf{x}),
  \label{eq:langevin_phi} \\
  \partial_t\pi(t, \mathbf{x}) &= -\frac{\delta H_\text{eff}}{\delta \phi}-\gamma\pi(t, \mathbf{x})+\xi(t, \mathbf{x}),
  \label{eq:langevin_pi}
\end{align}
where (after absorbing factors of $T$) $\phi$ is the effective field, $\pi$ is its canonical momentum, $\gamma$ is a damping parameter and $\xi$ is a local Gaussian noise term, satisfying
\begin{equation}
\langle \xi(t, \mathbf{x}) \xi(t', \mathbf{x}') \rangle = 2\gamma \delta(t-t')\delta^{(3)}(\mathbf{x}-\mathbf{x}').
\end{equation}
The effective Hamiltonian, $H_\text{eff}$, is 
constructed to match the long-wavelength equal-time correlation functions of the full quantum theory~\cite{Kajantie:1995dw, Braaten:1995cm}. This matching reduces the long-wavelength equilibrium thermodynamics of a four-dimensional (4d) quantum field theory to that of a 3d classical statistical field theory, and hence is known as high-temperature dimensional reduction (see Section~\ref{sec:dr} in the Supplemental Material for details).
Importantly, for a scalar field theory, an appropriate choice of $\gamma$ also leads to the matching of the unequal time correlation functions~\cite{Aarts:1997kp, Bodeker:1996wb,Aarts:1996qi}.

%%%%%%%%%%%%%%%%%%%%%%%%%%%%%%% Perturbative calculations %%%%%%%%%%%%%%%%%%%%%%%%%%%%%%%
 \section{Saddlepoint approximation}

Perturbative calculations of the bubble nucleation rate are based on saddlepoint approximations to the probability current between phases. To arrive at this from Eqs.~\eqref{eq:langevin_phi} and~\eqref{eq:langevin_pi}, one introduces the probability density $P(\phi,\pi)$, which satisfies a Fokker-Planck equation~\cite{Langer:1969bc, Berera:2019uyp}.
If the metastable state is sufficiently long-lived, the calculation can be set up as a static problem: one imposes boundary conditions that the metastable phase is populated thermally -- $P(\phi, \pi) \propto e^{-H_\text{eff}[\phi,\pi]}$ -- and the stable phase is unpopulated.
There is then a constant, small flux of probability over the barrier between phases, which does not appreciably deplete the metastable state nor appreciably populate the stable state~\cite{Kramers:1940zz}.

For field theories, the computation of this probability current was described by Langer~\cite{Langer:1967ax, Langer:1969bc}.
The relationship of Langer's formalism to high-temperature quantum field theories has been studied in Refs.~\cite{Gould:2021ccf, Hirvonen:2024rfg}, and generalised to higher orders in Ref.~\cite{Ekstedt:2022tqk}.
A crucial result is that the nucleation rate factorises
\begin{align}
\Gamma = A_\text{dyn} \times A_\text{stat},
\end{align}
where $A_\text{dyn}$ is the \textit{dynamical} factor and $A_\text{stat}$ is the \textit{statistical} factor. The latter is a purely time-independent quantity, equal to the vacuum nucleation rate in 2+1 dimensions. At leading (one-loop) order, $2\pi A_\text{dyn}$ is equal to the initial growth rate of the critical bubble.

The statistical part takes the form of a semiclassical path integral. Its computation requires first solving for a radially symmetric saddlepoint of the 3d tree-level action defined in Section~\ref{sec:dr} of the Supplemental Material, 
\begin{equation} \label{eq:bounce}
  \frac{\mathrm{d}^2\phi}{\mathrm{d} r^2} + \frac{2}{r} \frac{\mathrm{d} \phi}{\mathrm{d} r} = \frac{\mathrm{d}V_{3}(\phi)}{\mathrm{d}\phi}.
\end{equation}
The relevant solution, the critical bubble, has spherical symmetry. This reduces the problem to solving a one-dimensional boundary value problem, which be solved using the shooting method~\cite{Coleman:1977py}.

With the critical bubble $\phi_\text{b}$ in hand, as well as the position of the metastable phase $\phi_{0}$, the statistical part of the nucleation rate is proportional to the relative Boltzmann-weighted phase-space volume of these two stationary points,
\begin{align}
A_\text{stat} =
  \sqrt{\bigg|
    \frac{\det (S''[\phi_{0}]/2\pi)}
      {\det '(S''[\phi_\text{b}]/2\pi)}
  \bigg|}
  \left(\frac{\Delta S[\phi_\text{b}]}{2\pi}\right)^{3/2}
  e^{-\Delta S[\phi_\text{b}]} ,
\end{align}
where $S''$ denotes the second functional derivative of the action, the prime on $\text{det}'$ denotes that zero modes are not to be included, and $\Delta S[\phi_\text{b}]\equiv S[\phi_\text{b}] - S[\phi_{0}]$. The computation of the functional determinants can be carried out using the Gelfand-Yaglom theorem~\cite{Gelfand:1959nq, Baacke:1993ne}.

For the dynamical factor, one finds that
\begin{align}
A_\text{dyn} = \frac{1}{2\pi}\left(\sqrt{|\lambda_-|+\frac{\gamma^2}{4}} - \frac{\gamma}{2}\right),
\end{align}
where $\lambda_-$ is the single negative eigenvalue of $S''[\phi_\text{b}]$ and $\gamma$ is the damping rate of equation~\eqref{eq:langevin_pi}.

For the present theory, the computation of both $A_\text{stat}$ and $A_\text{dyn}$ can be carried out using the numerical package BubbleDet~\cite{Ekstedt:2023sqc}, or the numerical fits from Ref.~\cite{Ekstedt:2021kyx}. This yields our full one-loop rate, $\Gamma_\text{one-loop}$.

However, for applications of nucleation theory, especially for more complicated models, neither the functional determinant nor the dynamical prefactor are typically computed.
In Section~\ref{sec:lower_order_approximations} in the Supplemental Material we show two common lower-order approximations: in the first, $\Gamma_\text{tree-level}$, the functional determinant and dynamical prefactor are estimated as $T^4$, and in the second, $\Gamma_\text{LPA}$, a local potential approximation is made for the functional determinant, together with a prescription for dropping	 unwanted imaginary parts.

%%%%%%%%%%%%%%%%%%%%%%%%%%%%%%% Lattice simulations %%%%%%%%%%%%%%%%%%%%%%%%%%%%%%%
\section{Lattice simulations}
Our nonperturbative calculation of the nucleation rate closely follows the approach introduced in Refs.~\cite{Moore:2000jw,Moore:2001vf}.

The lattice discretisation of the action takes the form
\begin{align} \label{eq:lattice_action}
S_\text{lat} &= \sum_x a^3 \bigg[
- \frac{1}{2} Z_\phi \phi_x (\nabla_\text{lat}^2 \phi)_x
+ \sigma_\text{lat} \phi_x \nonumber
\nonumber \\&\quad
+ \frac{1}{2} Z_\phi Z_m m^2_\text{lat} \phi_x^2
+ \frac{1}{4!} Z_\phi^2 \lambda_\text{lat} \phi_x^4
\bigg],
\end{align}
where the sum over $x$ extends over the 3d lattice sites. We have removed the cubic term by a constant shift in the field. One can derive ex act lattice-continuum relations for this model within lattice perturbation theory~\cite{Laine:1995np}, and improvements up to $O(a^2)$~\cite{Arnold:2001ir, Moore:2001vf, Sun:2002cc}. See Section~\ref{sec:lattice_details} in the Supplemental Material for more information about our lattice action and~\cite{scalnuc} for the simulation code.

There are two stages: lattice Monte Carlo simulations which generate configurations, some of which lie close to the separatrix (sometimes known as the `transition surface'~\cite{Langer:1969bc, Gould:2021ccf}) of field configurations between the two phases.
Selected near-separatrix configurations are then evolved numerically in a thermal bath to determine whether they tunnel or not.

\subsection{Critical bubble probability}
\label{subsec:prob}

\begin{figure}
  \includegraphics*[width=0.95\columnwidth]{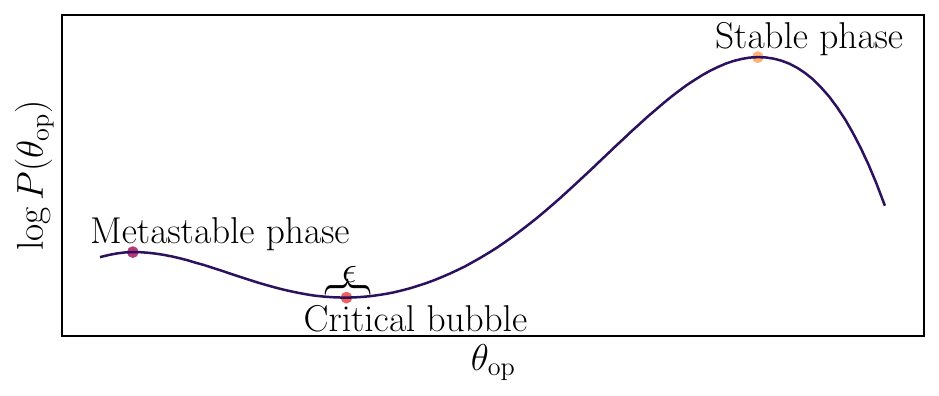}
  \caption{Probability distribution $P(\theta_\text{op})$ of the order parameter at some temperature below $T_\mathrm{c}$. The metastable and stable phases are separated by an exponentially suppressed area, the mixed critical bubble. The separatrix configurations are drawn from the narrow range $\epsilon$ around the critical bubble.}
  \label{fig:full_hist}
\end{figure}

The separatrix configurations are suppressed by $\sim e^{-65}$ at the temperature we simulate.
Our Monte Carlo simulations therefore use the multicanonical method~\cite{Berg:1991cf, Berg:1992qua} to overcome this suppression and generate a sufficient number of near-critical bubble configurations (see Ref.~\cite{Gould:2021dzl} for its application to the current model).
The order parameter $\op$ is measured, yielding a histogram that (below $T_c$) will resemble Fig.~\ref{fig:full_hist}. If the choice of order parameter, system volume and geometry admit, then there will be a local minimum between the metastable and stable phase peaks that we identify with the separatrix, at $\op = \crit$. This corresponds to critical bubble configurations. Configurations sufficiently near the separatrix $\op \in [\crit - \frac{\epsilon}{2}, \crit + \frac{\epsilon}{2}]$  are used as the initial conditions for evolution of the the lattice versions of the stochastic evolution equations~(\ref{eq:langevin_phi}) and~(\ref{eq:langevin_pi}), along with momenta drawn from a Gaussian distribution. The final result is independent of the exact choice of $\epsilon$, but it should be sufficiently small that the sampled configurations correspond to near-critical bubbles.

From the selected near-separatrix field configurations we can determine the probabilities~$P(\op)$ for the critical bubble and metastable phases and we normalise the probability of being in the critical configuration to that of the metastable phase,
\begin{equation}
    P_c^{\mathrm{normalised}}=\frac{P(|\op-\crit|<\epsilon/2)}{\epsilon P(\op<\crit)}.
\end{equation}

\subsection{Effective tunnelling fraction} \label{subsec:time}
\begin{figure*}
    \includegraphics[width=0.17\textwidth]{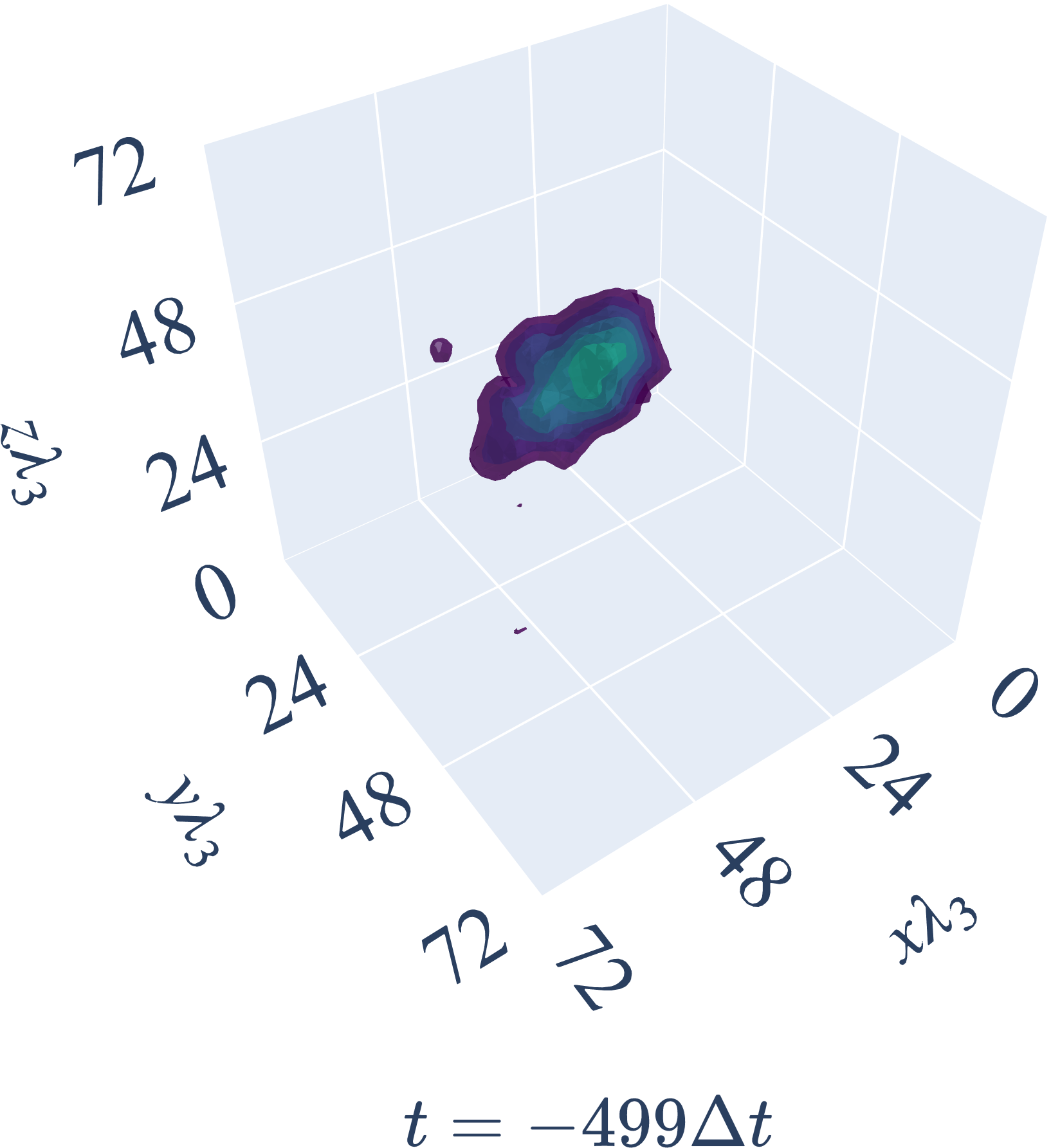}
    \hfill
    \includegraphics[width=0.17\textwidth]{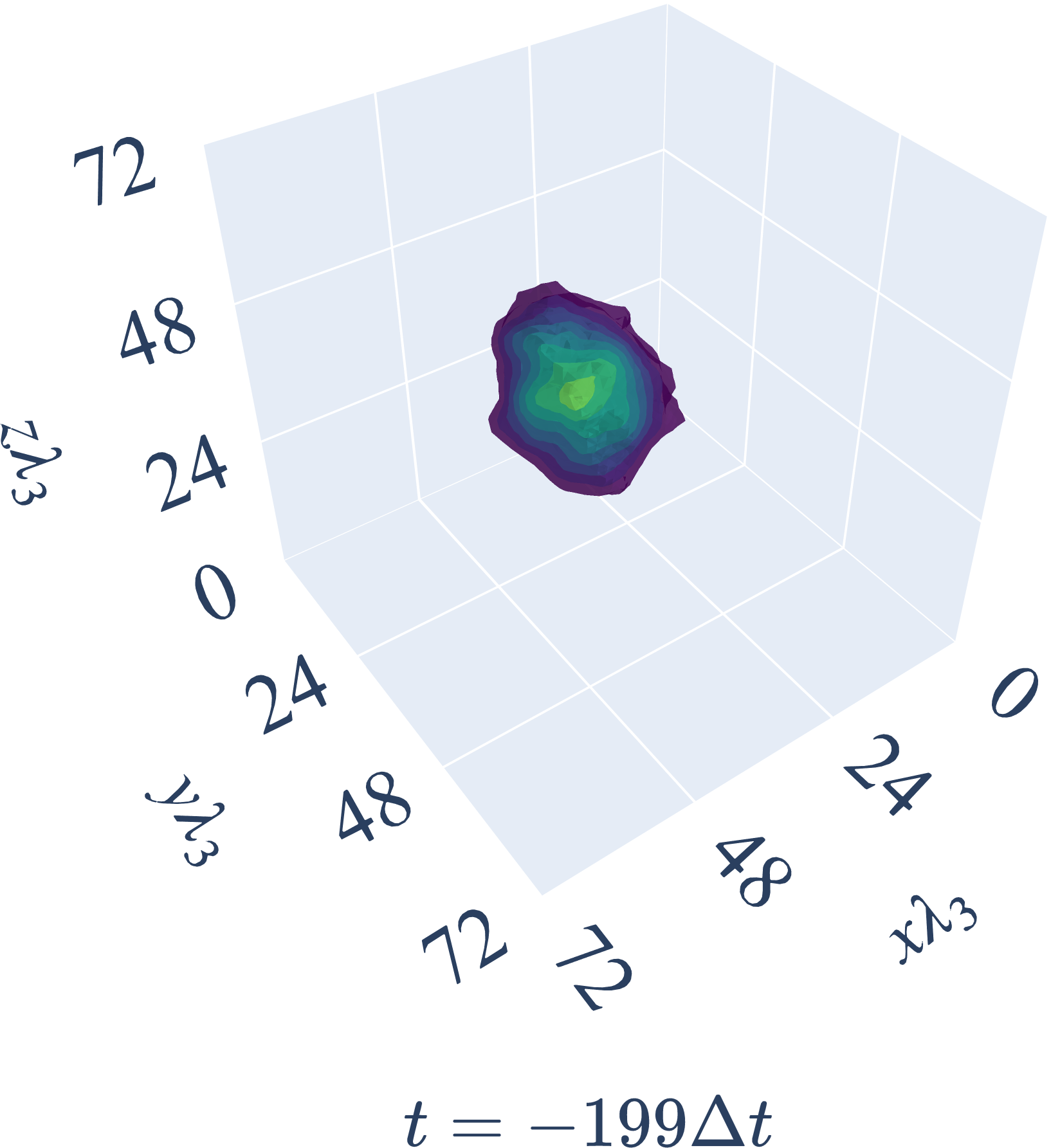}
    \hfill
    \includegraphics[width=0.17\textwidth]{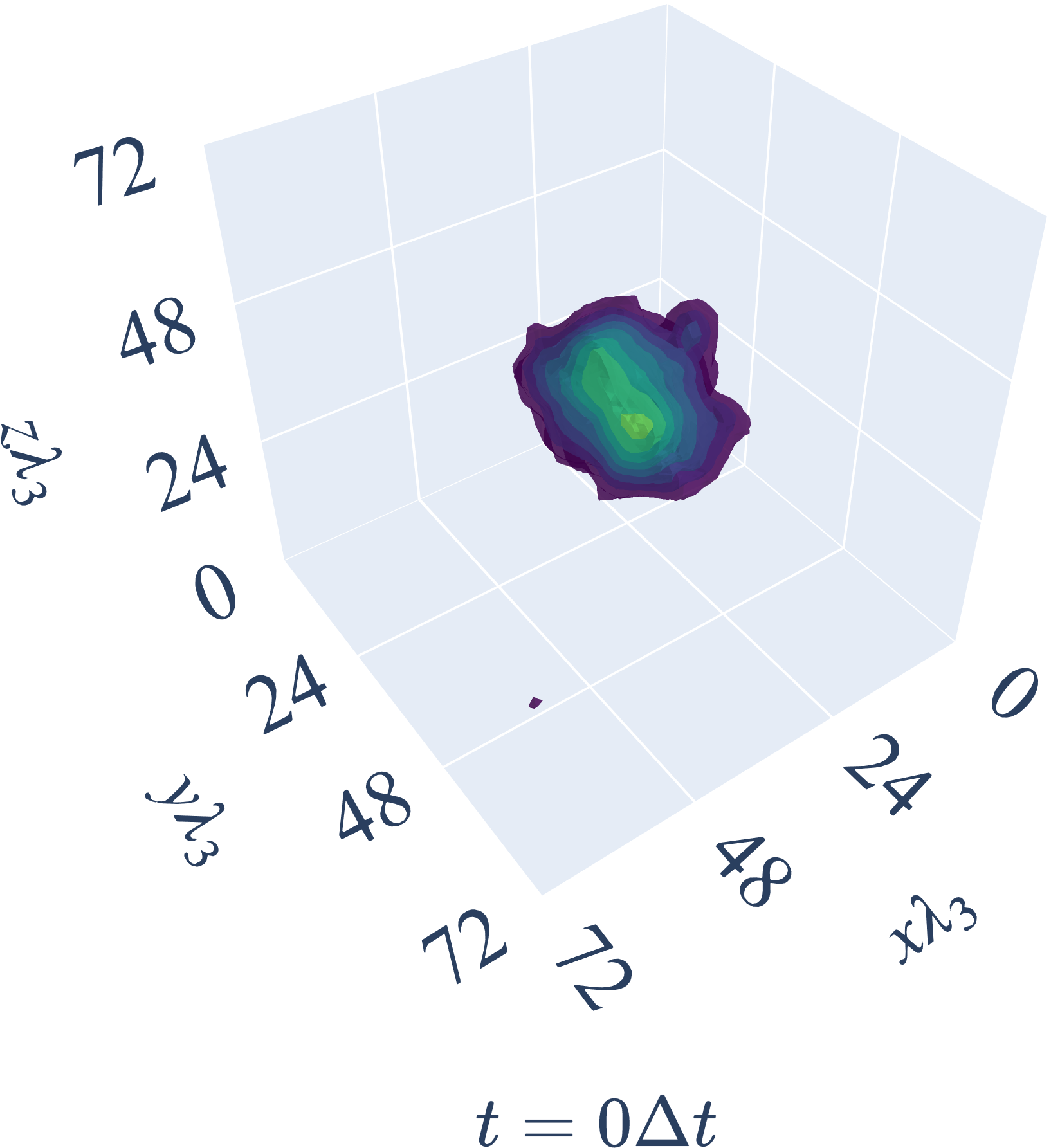}
    \hfill
    \includegraphics[width=0.17\textwidth]{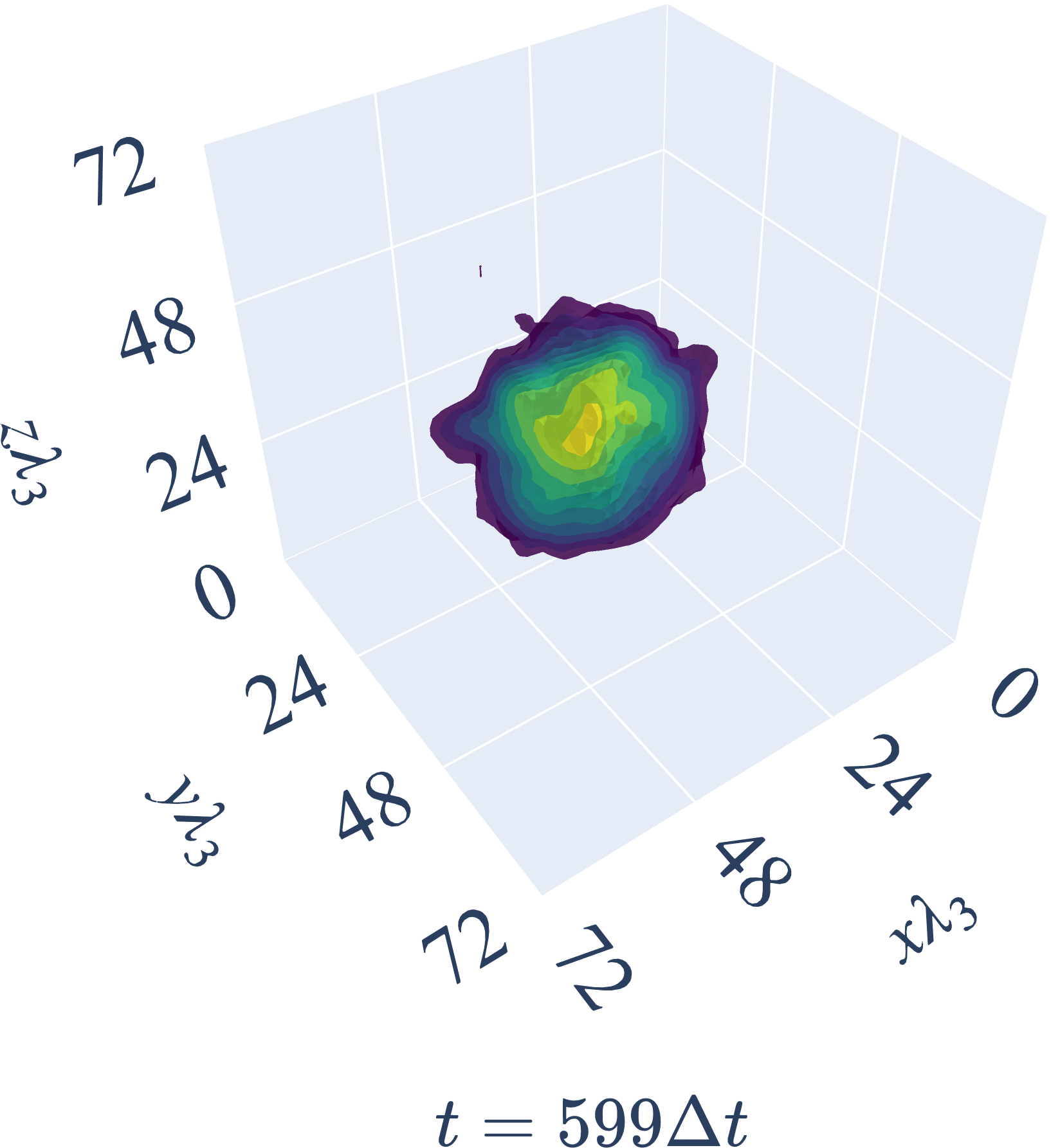}
    \hfill
    \includegraphics[width=0.17\textwidth]{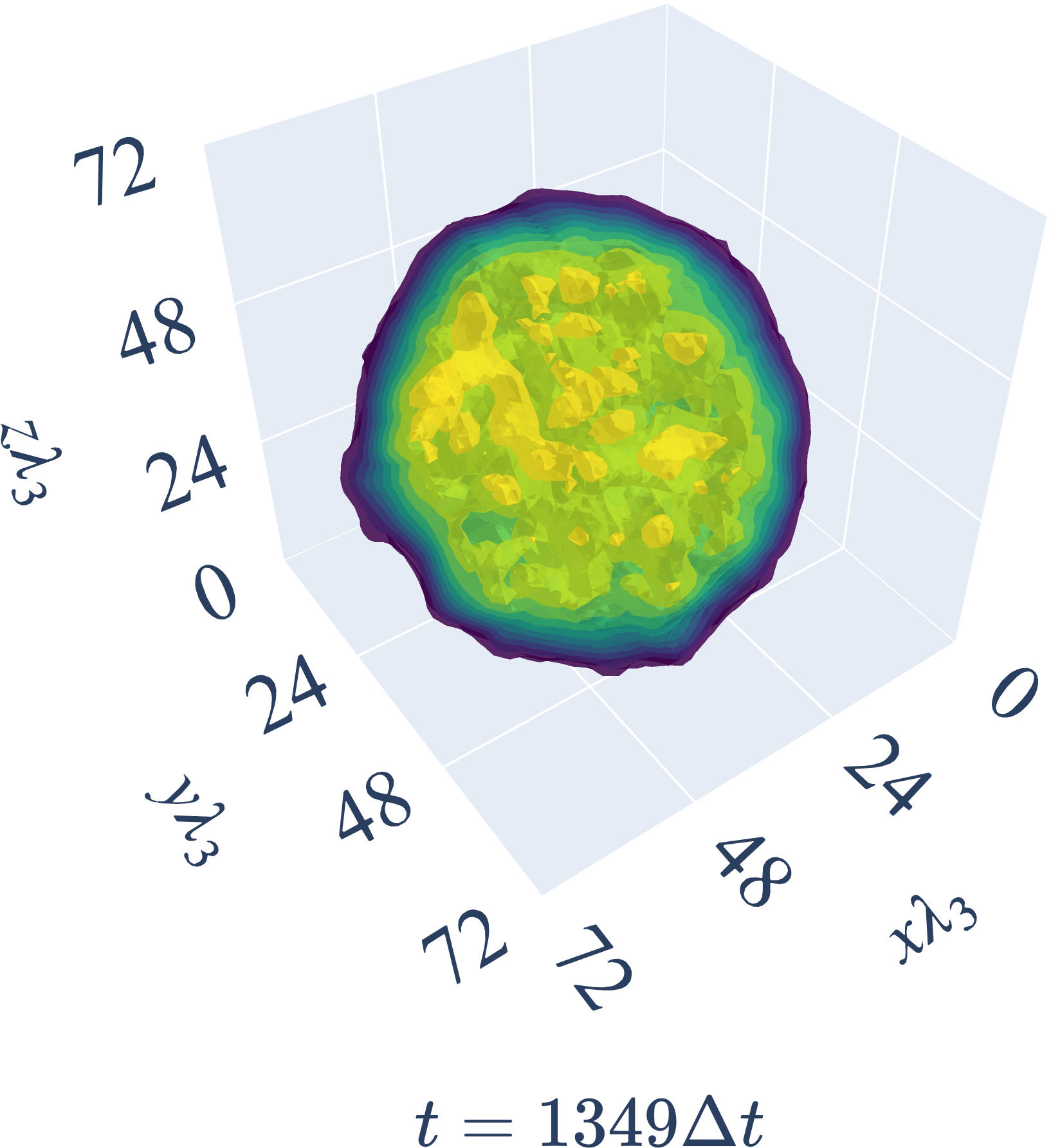}
    \hfill
    \includegraphics[width=0.04\textwidth]{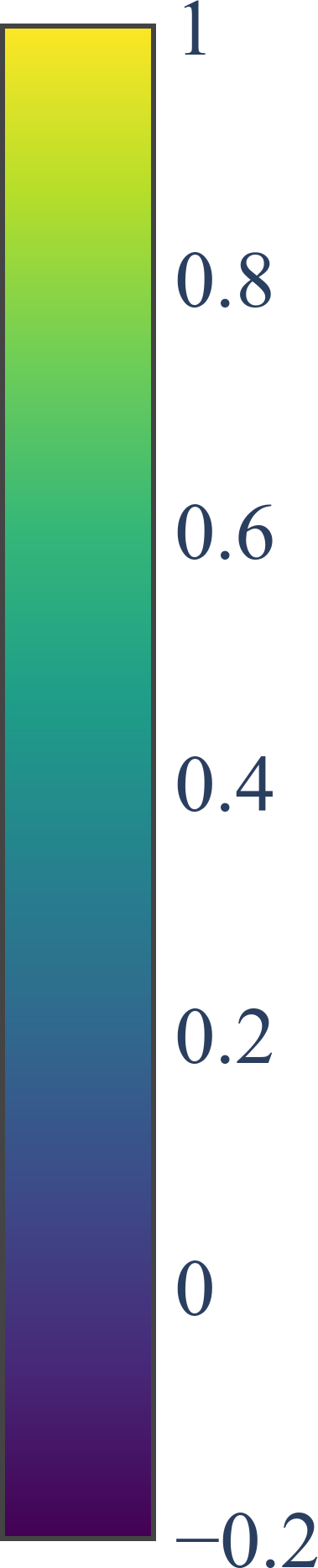}
  % }
    \caption{Snapshots of a nucleating bubble with lattice size $L\lambda_3 = 72$ and spacing $a\lambda_3=1.5$. 
    We remove ultraviolet fluctuations by performing eight steps of nearest neighbour averaging.
    Here, $t=0\Delta t$ corresponds to the initial separatrix configuration and the negative time direction describes the backward time evolution. Note that the snapshots are not evenly spaced in time.}
    \label{fig:visualisation}
\end{figure*}

Following Ref.~\cite{Moore:2001vf}, the dynamical information can be separated into two parts, the flux -- the rate of change of the order parameter as it crosses the separatrix -- and $\mathbf{d}$ -- the ratio of tunnelled configurations determined by real time trajectories. 
To determine whether a configuration tunnels or not, we directly evolve the stochastic Hamiltonian equations,~\eqref{eq:langevin_phi} and~\eqref{eq:langevin_pi}, both forwards and backwards in time, with the initial momenta reversed for the backwards evolution (see Fig.~\ref{fig:visualisation}). We use a timestep $\Delta t \lambda_3 = 0.01$ much smaller than the lattice spacing.

If a given trajectory begins and ends in different phases (either metastable~$\to$~stable or vice versa), we set $\delta_{\text{tunnel}}=1$, otherwise $\delta_{\text{tunnel}} = 0$. The ratio of tunnelled trajectories is an observable defined for each trajectory as
\begin{equation}
  \mathbf{d} = 
\frac{\delta_{\mathrm{tunnel}}}{N_{\mathrm{crossings}}},
\end{equation}
where $N_\mathrm{crossings}$ is the number of times the given trajectory crosses $\crit$.
Lastly, we must add a factor of $\frac{1}{2}$ to account for
metastable~$\to$~stable trajectories only, to avoid double counting the tunnelled trajectories.

The full nucleation rate on the lattice is a combination of the above three elements,
\begin{equation}
\Gamma \mathcal{V} = P_c^{\mathrm{normalised}}\left\langle \frac{1}{2} \mathrm{flux}\times \bf{d}\right\rangle,
\label{eq:unfactorised_rate}
\end{equation}
where $\mathcal{V}$ is the lattice volume.

For field-theory systems with many degrees of freedom, the flux is dominated by uncorrelated short-time fluctuations, whereas the global behaviour of the trajectories depends on longer-time correlations. These are therefore assumed to be uncorrelated and we approximate the rate by,
\begin{equation}
  \Gamma \mathcal{V} \approx P_c^{\mathrm{normalised}}\frac{1}{2} \left\langle \mathrm{flux}\right\rangle \left\langle \bf{d}\right\rangle.
\end{equation}

These formulae are order parameter independent,
but the choice of order parameter affects the viability of this method. At larger volumes, the contribution of bulk fluctuations about the metastable minimum will in general grow relative to the critical bubble. The physical volume occupied by the critical bubble is fixed, so that eventually bulk fluctuations dominate. 

The flux through the critical separatrix surface is order parameter dependent. It can be solved analytically due to the Gaussianity of the momentum field~\cite{Moore:2001vf}
\begin{equation}
  \langle\mathrm{flux}\rangle = \left< \left| \frac{\Delta \op}{\Delta t} \right|_{\crit} \right>
   = \sqrt{\frac{8}{\pi \mathcal{V}}(\crit+A^2)},
\end{equation}
where our order parameter $\op = \bar{\phi^2}-2A\bar{\phi}$ and $A$ is a constant which we choose to be the peak of the histogram of $\phi$ in metastable phase. We have used overbar to mean volume averaging and angle brackets for statistical averaging. We also considered  $\op' = \bar{\phi}$ (with corresponding flux $\sqrt{2/(\pi \mathcal{V})}$) but found that bulk fluctuations affected our ability to go to large volumes with this choice.

%%%%%%%%%%%%%%%%%%%%%%%%%%%%%%% Lattice resutls %%%%%%%%%%%%%%%%%%%%%%%%%%%%%%%
\subsection{Results} \label{sec:results}

\begin{figure}[tb]
  \centering
    \includegraphics[width=0.8\columnwidth]{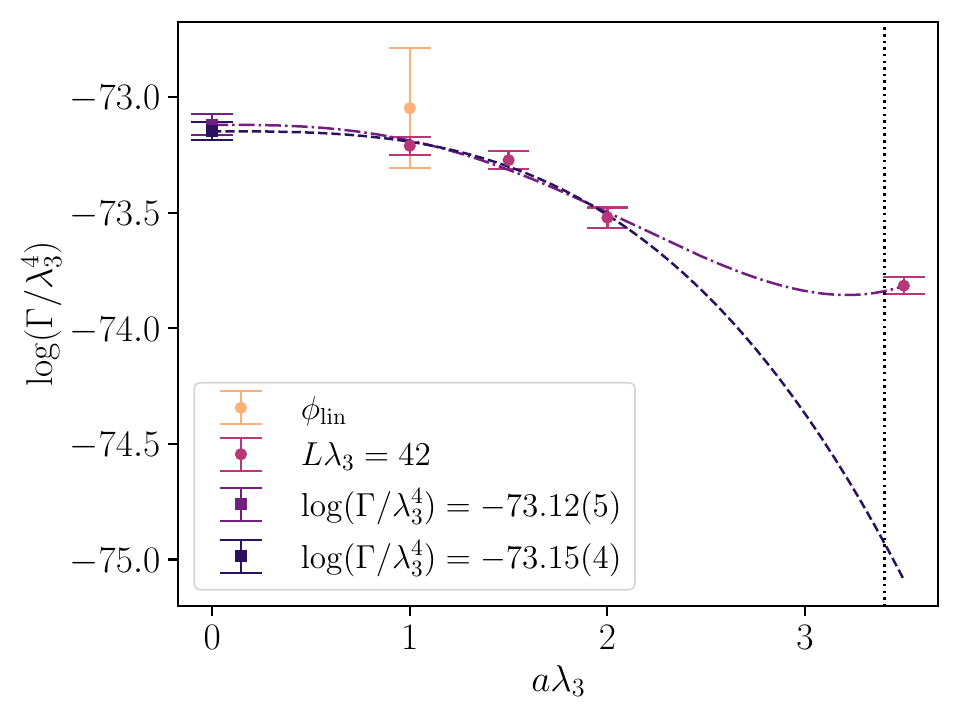}
    \\
    \includegraphics[width=0.8\columnwidth]{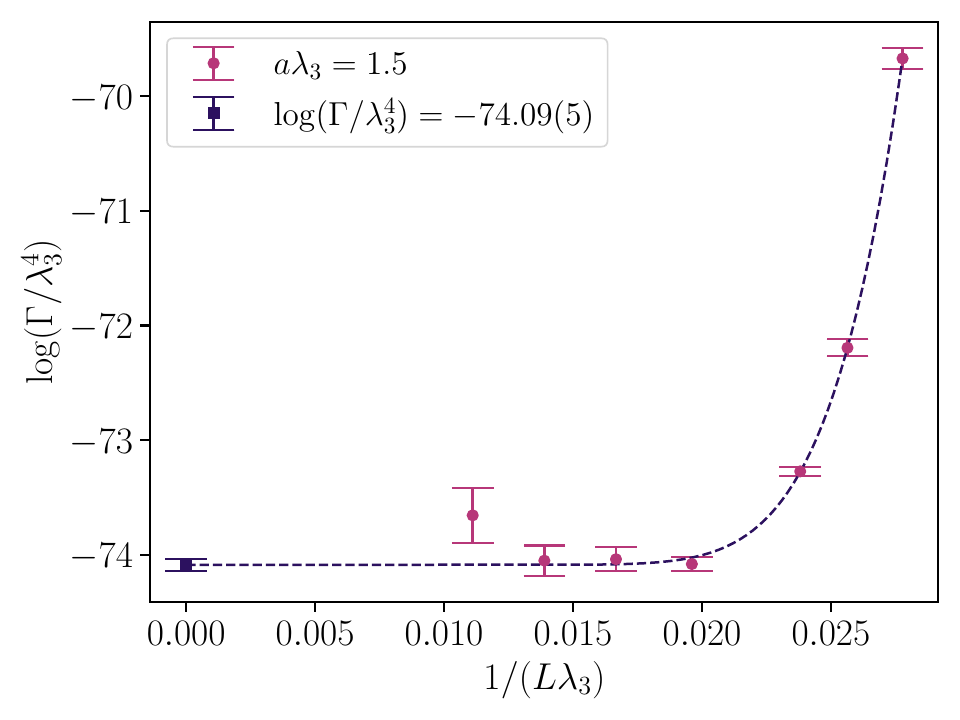}
    \caption{The zero lattice spacing (top) and infinite volume (bottom) extrapolations, together with cubic, cubic plus quartic and exponential fits respectively. The vertical line in the top plot marks a perturbative estimate of the correlation length.
    }
  \label{fig:limits}
  \end{figure}

In Figure~\ref{fig:limits}, we show how the obtained nucleation rate varies with lattice spacing and volume.
The dependence on both is rather mild, allowing us to take controlled continuum extrapolations.

Given our use of an $O(a^2)$ improved lattice discretisation, the leading dependence on lattice spacing arises at $O(a^3)$. For a fixed volume of $L\lambda_3 = 42$ we vary the lattice spacing and perform a least-squares fit $f(a)=b+c a^3$ to the logarithm of the rate. The best fit line, with $\chi^2/\mathrm{dof}=0.79$, excluding $\lambda_3 a = 3.5$, is plotted in the figure. Jackknife resampling this fit yields $\log (\Gamma/\lambda_3^4)=-73.24(11)$. Our largest lattice spacing $a\lambda_3 = 3.5$ is approximately the inverse screening mass of the system --- a one-loop estimate of the screening mass yields $m_\text{s}^\text{PT}/\lambda_3 = 0.294(1)$~\cite{Gould:2021dzl} --- at which point the cubic form is expected to break down. Motivated by this we fit $f(a)=b+ca^3 + da^4$ to all data points, giving $\chi^2/\mathrm{dof}=1.76$. We find agreement on the fit parameters between the two approaches. From the figure we deduce that lattice artefacts are comparable to statistical uncertainties already with $\lambda_3 a = 1.5$.

For reference we also show the nucleation rate computed using the linear order parameter $\op = \phi_\text{lin}$ for one lattice spacing. We find agreement within errors with our final results, however our quadratic order parameter shows substantially reduced errors, which we attribute to the relative suppression of the metastable phase bulk fluctuations.

We then show the dependence on volume at constant lattice spacing $\lambda_3 a = 1.5$. Away from the second-order phase transition, this 3d Euclidean model has no massless modes, so long-range correlations die off exponentially with distance. Motivated by this, we fit the logarithm of the rate to an exponential $f(L)=b + c \exp(-m_\text{s} L)$, finding $m_\text{s}/\lambda_3 = 0.283(12)$, consistent with the perturbative screening mass.
We show the best fit line in the figure, with $\chi^2/\mathrm{dof}=1.54$. Jackknife resampling this fit yields $\log (\Gamma/\lambda_3^4)=-74.09(5)$ for the infinite-volume extrapolation.
Note that fitting $f(L) = b+c/L^n$ with $n\in\{1,2,3\}$ to our finite-volume data yields much poorer fits with $\chi^2/\text{d.o.f.}\sim 10^2$.

Even with our improved choice of order parameter, we are limited in the largest volume we can reach for a number of reasons. For sufficiently large volumes, critical bubble configurations are again buried under the bulk fluctuations of the metastable phase.
Nevertheless, we see that the bubble is well resolved for the volumes we simulate in two ways: both geometrically determining which interface geometry (bubble, cylinder or slab) is favoured in the thin-wall approximation~\cite{Moore:2000jw}, and also visual inspection of configurations show a well-resolved spherical bubble configuration, see Fig.~\ref{fig:visualisation}.

  \begin{figure}[tb]
      \includegraphics[width=0.8\columnwidth]{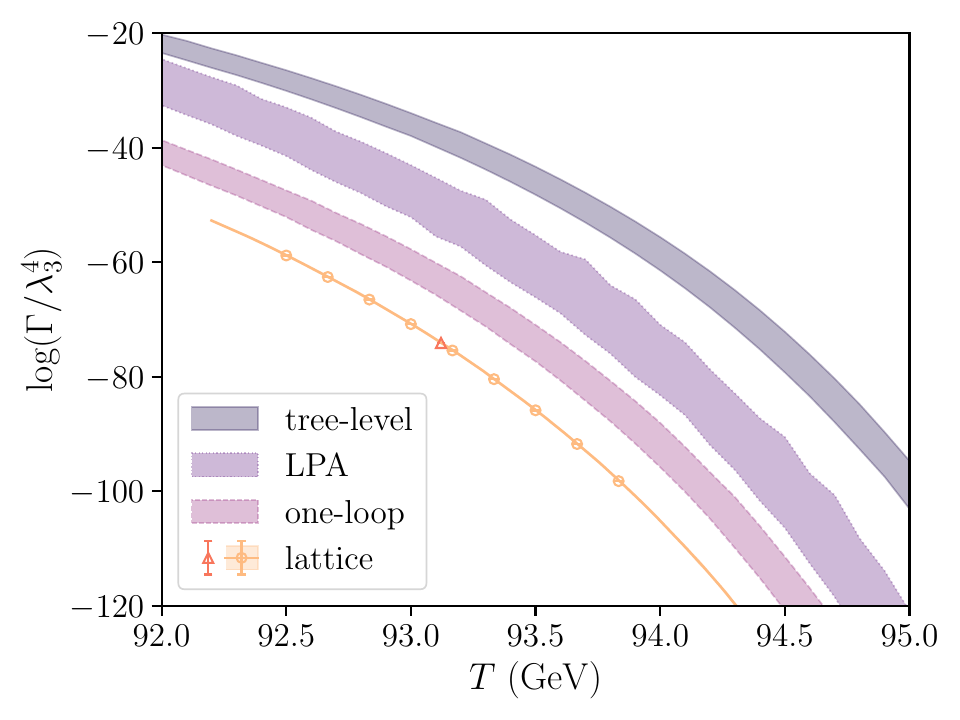}
      \caption{The nucleation rate as a function of temperature. Uncertainty
      bands for the tree-level and one-loop perturbative results are based on
      varying the renormalisation scale over $\mu_3/\lambda_3 \in \{0.5, 1,
      2\}$. The uncertainty estimate on the LPA relflects different choices for
      removing imaginary parts of the potential. The lattice points are
      continuum extrapolated; the red triangle highlights the temperature
      actually simulated, while the orange circles have utilised reweighting.
      The orange continuous line is the reweighted result for $a\lambda_3=1.5$,
      $L\lambda_3=60$. The results in this figure are tabulated at~\cite{gould_2024_11085693}.%
      \footnote{The one-loop result in this figure is corrected in the erratum
      at the end of this document.}}
      \label{fig:rate}
  \end{figure}

%%%%%%%%%%%%%%%%%%%%%%%%%%%%%%% Discussion %%%%%%%%%%%%%%%%%%%%%%%%%%%%%%%
\section{Discussion}

The main result of this work is the logarithm of the nucleation rate, reliably computed to high precision for a real scalar quantum field theory at high temperature. Statistical uncertainties are much smaller than discrepancies with other methods, as seen in Fig.~\ref{fig:rate}. Systematic uncertainties related to the continuum extrapolation are also well under control, as evidenced by Fig.~\ref{fig:limits}. This has been made possible by use of efficient multicanonical algorithms~\cite{Berg:1991cf, Berg:1992qua, Moore:2000jw, Moore:2001vf}, and exact and $O(a^2)$ improved lattice-continuum relations~\cite{Laine:1995np, Arnold:2001ir, Moore:2001vf, Sun:2002cc}.

The benchmark parameter point that we simulated was chosen with the hope of testing perturbation theory in a regime where it is expected to work well.%
\footnote{For our chosen xSM parameter point in 4d, at one-loop we find $T_\text{c}\approx 98.5~\text{GeV}$, $\Delta\langle \varphi \rangle_\text{c}/T_\text{c}\approx 1.67$ and the percolation temperature is $T_{*}\approx94.8~\text{GeV}$~\cite{Caprini:2019egz}.}	
The loop expansion converges rapidly for the field condensate (adopting units where $\lambda_3=1$)~\cite{Gould:2021dzl}

\begin{equation} 
  \Delta\langle \phi \rangle_\text{c}^\text{lat}=1.341(2),
  \quad
  \Delta\langle \phi \rangle_\text{c}^\text{pt}=1.2+0.1378+\dots .
\end{equation}
We have displayed the first two terms of the loop expansion within the 3d EFT. The tree-level prediction agrees with the lattice to 10\%, and at one-loop to 0.2\%. The two- and three-loop terms are smaller still. In Sec.~\ref{sec:results}, we have also found that lattice and one-loop estimates of the screening mass agree within statistical uncertainties.

Yet for the nucleation rate at the temperature simulated, and at the same
physical parameter point, we have%
\footnote{
  The dynamical factor was missing in previous versions, which modifies
  this equation at one-loop;
  see the erratum at the end of this paper.
}
\begin{equation} \label{eq:rate_numbers_compared}
  |\log\Gamma|^\text{lat} =74.09(5),
  \quad
  |\log\Gamma|^\text{pt} =38.0+25.3+\dots
\end{equation}

One might expect that this lack of convergence could be
explained by the growth of the loop expansion parameter as the system supercools, which becomes almost twice as large at the temperature we have simulated than at the critical temperature.
However, even at twice larger couplings the discontinuity in the field condensate agrees to 20\% at tree-level and 2\% at one-loop~\cite{Gould:2021dzl}.
Note that the discrepancy in
the rate
cannot be explained by an offset in
temperature,
as both lattice and perturbation theory use the same EFT parameters, and
in this model the critical temperature is fixed by the $Z_2$ symmetry between phases, and hence is exact already at tree-level within the 3d EFT~\cite{Gould:2021dzl}.

However, what if the discrepancy between the one-loop and lattice results cannot be explained simply by the next loop order? While the critical bubble is a saddlepoint of the path integral, there may be other relevant saddlepoints~\cite{Andreassen:2016cvx, Braden:2018tky}. Further, for such weakly-damped systems, it has been argued that the saddlepoint approximation itself breaks down for the dynamical part of the nucleation rate~\cite{Hanggi:1990zz,Ekstedt:2022tqk}. Alternatively, a number of suggestions have been put forward to resolve the longstanding discrepancy between theory and experiment for the nucleation rate between the A and B phases of superfluid $^3$He, including resonant tunnelling and tunnelling via intermediate solitonic configurations~\cite{Hong:1987ur, Tye:2011xp}.

As far as we are aware, there has never been a calculation of the thermal nucleation rate of any quantum field theory beyond one-loop order.
To really test whether perturbation theory provides a reliable guide to bubble nucleation, a complete two-loop calculation at our benchmark point provides a clear target for future work. Further, nonperturbative results over a range of couplings could test convergence.

For phenomenological studies of cosmological phase transitions, and their gravitational wave signals, our results demonstrate that the widely-used tree-level and local potential approximations give relatively poor accuracy for the logarithm of the bubble nucleation rate. The full one-loop approximation fares significantly better. This calculation can be carried out in simple models using existing numerical tools~\cite{Ekstedt:2023sqc}, but is still out of reach for e.g.~scalar extensions of the Standard Model.

While our lattice simulations give reliable and precise predictions for the nucleation rate, their calculation is slow. To use such lattice simulations for phenomenological studies of cosmological phase transitions, faster sampling of the relevant bubble configurations is needed. We have identified one direction of progress, through the development of order parameters optimised to enhance the sampling of tunnelling configurations. Our quadratic order parameter $\op = \bar{\phi}^2 - 2A \bar{\phi}$ was crucial for efficiently simulating the larger volumes. 

\begin{acknowledgments}
  The authors would like to thank Kari Rummukainen for his insights and advice at a number of points throughout this study.
  O.G. (ORCID ID 0000-0002-7815-3379) was
  supported by
  the Research Funds of the University of Helsinki,
  U.K. Science and Technology Facilities Council (STFC) Consolidated Grant
  ST/T000732/1,
  a Research Leadership Award from the Leverhulme Trust,
  and a Royal Society Dorothy Hodgkin Fellowship.
  A.K. (ORCID ID 0000-0002-0309-3471) was supported by Research Council of Finland grant no. 328958 and a travel grant from the Jenny and Antti Wihuri Foundation.
  D.J.W. (ORCID ID 0000-0001-6986-0517) was
  supported by Research Council of Finland grant nos. 324882, 328958, 349865 and 353131.
  The authors also wish to acknowledge CSC -- IT Center for Science, Finland, for computational resources. 
\end{acknowledgments}

\section*{Data access statement}
Data supporting this paper is available from Zenodo at \url{https://doi.org/10.5281/zenodo.10891523}.

\bibliography{scalnuc_short}

\clearpage

\begin{widetext}
\begin{center}

  {\large \bf Supplemental material}

\end{center}
\end{widetext}

\setcounter{section}{0}

%%%%%%%%%%%%%%%%%%%%%%%%%%%%%%% Dimensional reduction %%%%%%%%%%%%%%%%%%%%%%%%%%%%%%%
\section{High-temperature effective field theory}
\label{sec:dr}

At high temperatures, the equilibrium thermodynamics of a weakly coupled 3+1 dimensional quantum field theory is captured by a 3 dimensional EFT containing only the light bosonic fields~\cite{Kajantie:1995dw, Braaten:1995cm}. The parameters of this EFT depend on those of the full theory and on the temperature. The precise matching relations can be computed order-by-order in powers of couplings.

In the vicinity of a phase transition, the effective mass of the field undergoing the transition becomes small, and hence the longest wavelength modes may often be described by an EFT for this field alone, with all other fields integrated out. We will assume that this is the case for the real scalar field, in which case the effective 3d Lagrangian is
\begin{align}
  \mathscr{L}_\text{eff} & = \frac{1}{2}\partial_i \phi \partial_i \phi + V_3(\phi), \\
  \text{where} \quad V_3(\phi) & = \sigma_3 \phi + \frac{m_3^2}{2} \phi^2 + \frac{g_3}{3!} \phi^3 + \frac{\lambda_3}{4!} \phi^4,
\end{align}
where $i=1,2,3$ runs over the spatial dimensions. Note that the field has mass dimension 1/2, being canonically normalised in 3 dimensions. To leading order, it corresponds to the zero Matsubara mode of the 4d field $\varphi$ divided by $\sqrt{T}$.

For the real-scalar extended Standard Model, the parameters of the effective action are, at leading order,
\begin{align} \label{eq:dr_start}
\sigma_3(T) &= \frac{\sigma}{\sqrt{T}} + \frac{1}{24}(g + 4 \kappa_1)T^{3/2}, \\
m_3^2(T) &= m^2 + \frac{1}{24}(\lambda + 4 \kappa_2)T^2, \\
g_3(T) &= \sqrt{T} g, \\
\lambda_3(T) &= T \lambda. \label{eq:dr_end}
\end{align}
At this leading order, the dimensional reduction relations take the same form for a wide range of interactions to the real scalar. For example, coupling to a Dirac fermion through a Yukawa interaction, $J_1=y \bar{\psi} \psi$, instead of to the Standard Model Higgs, would simply replace $\kappa_1 \to y m_\psi$, $\kappa_2 \to y^2$ in Eqs.~\eqref{eq:dr_start} to \eqref{eq:dr_end}, where $m_\psi$ is the tree-level mass of the fermion~\cite{Gould:2021ccf}.

High-temperature dimensional reduction describes only the equilibrium thermodynamics of our model. The real-time dynamics of $\phi$ are described by the Langevin equations given in equation~\eqref{eq:langevin_phi} and \eqref{eq:langevin_pi}~\cite{Aarts:1996qi, Aarts:1997kp, Bodeker:1996wb, Greiner:1996dx}. The relevant effective Hamiltonian is
\begin{equation}
  H_\text{eff} = \int \mathrm{d}^3 x \left[
    \frac{1}{2}\pi^2 + \frac{1}{2}\partial_i \phi \partial_i \phi + V_3(\phi)
  \right].
\end{equation}

The parameter $\gamma$ in equation~\eqref{eq:langevin_pi} describes the damping and fluctuations that $\phi$ experiences due to hard thermal fluctuations. This parameter can be found by ensuring that the Langevin equation reproduces the long-wavelength real-time correlation functions of the underlying quantum field theory. The result is that $\gamma=O(\lambda^{3/2} T/\pi)$~\cite{Greiner:1996dx}, so that the damping and noise terms are subdominant in equation~\eqref{eq:langevin_pi}. At leading order $\gamma$ may therefore be taken to be zero.

However, a nonzero positive value for $\gamma$ can help reduce finite size effects due to the heating of the lattice upon bubble nucleation. Thus, in our simulations we follow Ref.~\cite{Moore:2001vf} and take $\gamma=1/L$, thereby reaching $\gamma \to 0_+$ in the infinite volume limit.

For our benchmark point of the xSM, we have chosen
\begin{align*}
  M_\phi &= 240~\text{GeV}, & \sin\theta &= 0.1, & \kappa_2 &= 1.5,\\
  g &= -223.75~\text{GeV}, & \lambda &= 1.5489,
\end{align*}
where $M_\phi$ is the physical scalar mass, and $\theta$ the Higgs mixing angle, both evaluated at tree-level; for details, see Appendix A of Ref.~\cite{Gould:2019qek}. The corresponding Lagrangian parameters are $\sigma=-5.1340\times 10^5~\text{GeV}^3$, $\kappa_1 = 16.937~\text{GeV}$ and $m=108.23~\text{GeV}$. 

At the critical temperature (in this case $T_\text{c}=98.513~\text{GeV}$), this goes through the second benchmark point studied in Ref.~\cite{Gould:2021dzl}. The critical temperature is exact within the 3d EFT, as it is protected by a $Z_2$ symmetry. The equilibrium thermodynamics at the critical temperature was found to be under good perturbative control at one-loop order, at which point the dimensionless loop-expansion parameter within the EFT is $\alpha_3\approx 0.13$~\cite{Gould:2021dzl}. The temperature of spinodal decomposition, where the mass in the metastable phase goes through zero, is $T_\text{s}=89.920~\text{GeV}$ at tree-level, though this is corrected by loops within the EFT.

Our lattice simulations are carried out within the 3d EFT. The majority of our simulations were carried out at the parameter point corresponding to $T = 93.121~\text{GeV}$, where we have used Eqs.~\eqref{eq:dr_start} to~\eqref{eq:dr_end} to fix the corresponding 3d effective parameters. The dimensionless loop-expansion parameter at this temperature is $\alpha_3\approx 0.23$~\cite{Gould:2021dzl}. Note that, while we have used only leading order dimensional reduction matching relations, our calculations within the 3d EFT are valid nonperturbatively at the values of the effective parameters studied.

\begin{table}[h]
  \begin{ruledtabular} % makes it look nice and revtex'y
  \begin{tabular}{cccccc} % vertical lines are not revtex'y
  % \toprule
  $T$ (GeV) & $\mu_3 / \lambda_3$ & $\sigma_3/\lambda_3^{5/2}$ & $m_3^2/\lambda_3^2$ & $g_3/\lambda_3^{3/2}$ & $\lambda_3$ (GeV) \\
  \hline
  \rule{0pt}{3ex} % add some space
  93.121 & 1 & $-0.016687$ & $-0.082770$ & 0 & 144.23
  % \bottomrule
  \end{tabular}
\end{ruledtabular}
  \caption{Parameters of our lattice simulations. Note that neither $T$ nor $\lambda_3$ explicitly enter the simulations: the temperature is used to fix the 3d parameters through dimensional reduction, and $\lambda_3$ can be scaled out by dimensional analysis.}
  \label{tab:physicalparams}
\end{table}

%%%%%%%%%%%%%%%%%%%%%%%%%%%%%%% Semiclassical approximations %%%%%%%%%%%%%%%%%%%%%%%%%%%%%%%
\section{Lower order semiclassical approximations}
\label{sec:lower_order_approximations}

A common approximation to the nucleation rate is to take
\begin{align} \label{eq:gamma_tree}
  \Gamma_\text{tree-level} = T^4 \left(\frac{\Delta S[\phi_\text{b}]}{2\pi}\right)^{3/2} e^{-\Delta S[\phi_\text{b}]},
\end{align}
where $T^4$ replaces the functional determinant and dynamical prefactor.

The local potential approximation (LPA) is an alternative and widely-used approximation, utilising the one-loop effective potential when solving for the critical bubble,
\begin{align}
  V_\text{one-loop}(\phi) &= -\frac{1}{12\pi}V_{3}''(\phi)^{3/2}.
\end{align}
This is imaginary where $V_{3}''(\phi)<0$, such as on the tree-level potential barrier between phases. Such field values are absolutely unstable (as opposed to metastable), and the imaginary part of the potential gives the corresponding decay rate, which, unlike bubble nucleation, is not exponentially suppressed~\cite{Weinberg:1987vp}. The imaginary part arises because the computation of the effective potential assumes the background field is constant, yet for the critical bubble this assumption fails~\cite{Croon:2020cgk, Gould:2021ccf}.
To bypass this complication, one throws away the imaginary part in some ad-hoc way, for example one of
\begin{align}  \label{eq:adhoc_real_potential}
V_\text{LPA}(\phi) = 
V_{3}(\phi) + 
\begin{cases}
   \frac{-1}{12\pi}|V_{3}''(\phi)|^{3/2},\\
   \frac{-1}{12\pi} \text{Re}\left(V_{3}''(\phi)^{3/2}\right).
\end{cases}
\end{align}
Note that these different choices will yield different physical results. The former choice~\cite{Wainwright:2011kj} is negative for $V_{3}''(\phi)<0$, while the latter choice~\cite{Delaunay:2007wb, Athron:2022jyi} is zero. One then solves equation~\eqref{eq:bounce} using $V_\text{LPA}(\phi)$ as the potential, and inserts the resulting action into equation~\eqref{eq:gamma_tree} to yield the LPA rate, $\Gamma_\text{LPA}$.

%%%%%%%%%%%%%%%%%%%%%%%%%%%%%%% Lattice details%%%%%%%%%%%%%%%%%%%%%%%%%%%%%%%
\section{Lattice details}
\label{sec:lattice_details}

Here we give further details of our lattice discretisation. We outline the purely spatial action used in the Monte-Carlo simulations, as well as details of our real-time update algorithm. Our simulation code is available at~\cite{scalnuc}.

For the discretised action we use, Eq~\eqref{eq:lattice_action}, possible discretisations of the lattice Laplacian include
\begin{align}
(\nabla_\text{lat}^2 \phi)_x^{(1)} &= \sum_i \frac{1}{a^2}\left(\phi_{x+i} + \phi_{x-i} - 2\phi_x\right), \label{eq:laplacian_Oa2}\\
(\nabla_\text{lat}^2 \phi)_x^{(2)} &= \sum_i \frac{1}{a^2}\bigg(
-\frac{1}{12}\phi_{x+2i}
+ \frac{4}{3}\phi_{x+i}
\nonumber \\&\quad
- \frac{5}{2}\phi_x
+ \frac{4}{3}\phi_{x-i}
-\frac{1}{12}\phi_{x-2i} \bigg), \label{eq:laplacian_Oa4}
\end{align}
where $i$ runs over the three directions of the lattice. These expressions have $O(a^2)$ and $O(a^4)$ errors accurate respectively. For smooth field configurations, such as one encounters in classical field theory in the absence of fluctuations, one can simply set the lattice parameters equal to their renormalised continuum counterparts. However, in the presence of thermal fluctuations the lattice action requires renormalisation, and the relationships between lattice and renormalised continuum coefficients are modified as $\kappa_\text{lat} = \kappa_\text{\MSbar} + \delta\kappa$.

The interaction terms of this 3d real scalar theory all have coefficients with positive mass dimension. As a consequence the way that physical quantities can depend on the lattice spacing $a$ is tied to the way that they depend on the couplings. In the approach to the continuum limit, the leading $a$ dependence is determined by dimensional analysis up to a small number of constants, which can be computed in lattice perturbation theory. Further, the lattice regularisation scheme can be related to other regularisation schemes, such as \MSbar, by matching perturbative computations of the vacuum energy~\cite{Laine:1995np, Laine:1997dy}.

For the real scalar theory, shifting the field by a constant so that $g_\text{lat}=0$ ensures that the coefficients of odd powers of the field are not renormalised, i.e.~$\delta \sigma_3 = 0$ and  $\delta g_3 = 0$. One can then extract the lattice-continuum relations from Refs.~\cite{Arnold:2001ir, Moore:2001vf, Sun:2002cc},
\begin{align}
\label{eq:lattice_continuum_start}
\delta m_3^2  &= -\frac{\Sigma \lambda_3}{2(4\pi) a}
+ \frac{\lambda_3^2}{(4\pi)^2}\left[\log\frac{6}{a\mu_3} + C_3 - \Sigma\xi\right], \\
\delta \lambda_3 &= \frac{3\xi \lambda_3^2 a}{2(4\pi)}
+ \frac{\lambda_3^3 a^2}{(4\pi)^3}\left(\frac{3}{4}\xi^2 - 3 C_1 - \frac{C_2}{3}\right), \\
Z_\phi &= 1
+ \frac{C_2 \lambda_3^2 a^2}{6(4\pi)^2}, \\
Z_{m} &= 1
+ \frac{\xi \lambda_3 a}{2(4\pi)}
+ \frac{\lambda_3^2a^2}{(4\pi)^2}\left(\frac{\xi^2}{4} - \frac{C_1}{2} - \frac{C_2}{6}\right),
\label{eq:lattice_continuum_end}
\end{align}
which all receive corrections at $O(a^3)$, except the squared mass parameter, which receives corrections already at $O(a)$.
In these equations $\Sigma$, $\xi$, $C_1$, $C_2$ and $C_3$ are numerical constants which were computed in Ref.~\cite{Moore:2001vf}. Their values depend on the scalar propagator, and hence on the choice of lattice Laplacian. For the $O(a^4)$ accurate Laplacian, which we have used for our final results, their values are
\begin{align}
\Sigma &= 2.75238391130752,\\
\xi &= -0.083647053040968, \\
C_1 &= 0.0550612,\\
C_2 &= 0.0334416,\\
C_3 &= -0.86147916.
\end{align}
Note that if the $O(a^2)$ terms in Eqs.~\eqref{eq:lattice_continuum_start}-\eqref{eq:lattice_continuum_end} are retained, one must use the higher order accurate Laplacian, Eq.~\eqref{eq:laplacian_Oa4} for the kinetic term to have the same accuracy.

Eqs.~\eqref{eq:lattice_continuum_start}-\eqref{eq:lattice_continuum_end} give the relationship between bare and \MSbar\ parameters up to cubic $O(a^3)$ corrections, except for the mass, which still has linear $O(a)$ corrections. However, the residual lattice-spacing dependence of the mass squared is of the form $c_3 \lambda_3^3 a + c_4 \lambda_3^4 a^2$ with $c_3$ and $c_4$ pure dimensionless numbers. These corrrections are independent of the mass. As a consequence, they cancel when considering differences such as $m_3^2 - m_{3,\text{c}}^2$, a trick which can be used to ensure faster convergence to the continuum limit for physical quantities.

The real time evolution is a mix of the Forest-Ruth algorithm~\cite{Forest:1989ez} and momentum refresh (sometimes referred as partial momentum refreshment or partial momentum Monte Carlo)~\cite{Horowitz:1991rr,Moore:2001vf, Neal:2011mrf}. Forest-Ruth is a symplectic and fourth order accurate algorithm, which can be constructed starting from the second order accurate leapfrog aglorithm~\cite{Yoshida:1990zz}. In our case, in order to make measurements at every timestep, we utilise the kick-drift-kick form of the leapfrog algorithm. For the field $\phi_{t,x}$ and momentum $\pi_{t,x}$ this latter update algorithm is
\begin{align}
  \pi_{t+\frac{1}{2},x} &=\pi_{t,x} - \frac{1}{a^3}\frac{\partial H_\text{eff}}{\partial \phi_{t,x}}\frac{\Delta t}{2} ,\\
  \phi_{t+1,x} &= \phi_{t,x} + \pi_{t+\frac{1}{2},x}\Delta t ,\\
  \pi_{t+1,x} &=\pi_{t+\frac{1}{2},x} - \frac{1}{a^3}\frac{\partial H_\text{eff}}{\partial \phi_{t+1,x}}\frac{\Delta t}{2} .
\end{align}
Forest-Ruth then consists of three successive kick-drift-kick updates, with timesteps, $\Delta t_1$, $\Delta t_2$ and $\Delta t_3$, given by
\begin{align}
  \Delta t_1 &= (2-2^{1/3})^{-1} \Delta t ,
  \\
  \Delta t_2 &= -2^{1/3}(2-2^{1/3})^{-1} \Delta t, 
  \\
  \Delta t_3 &= \Delta t_1 . 
\end{align}
Being symplectic, for $\gamma = 0$ this algorithm conserves energy for long times, with errors of order $O(\Delta t /a)^5$. For our choice of timestep $\Delta t \lambda_3 = 0.01$, we have demonstrated energy convergence at the level of $O(10^{-6})$. We have also demonstrated that for $\gamma = 0$ the algorithm is reversible to numerical precision.

In order to reproduce the damping $\gamma$ and Gaussian random noise $\xi$ of the Langevin Eqs.~\eqref{eq:langevin_phi} and~\eqref{eq:langevin_pi}, we use the momentum refresh. In our case the update reads
\begin{align}
  \pi_{t+0,x} &= \sqrt{1-\vartheta^2}\pi_{t-0,x} + \vartheta \xi_{t,x} ,\\
  \vartheta^2 &= 1-\exp(-2\gamma\Delta t) .
\end{align}
The momentum refresh is applied after each Forest-Ruth evolution step, together completing one full time evolution iteration.

This complete algorithm conserves the thermal distribution of the momenta with
errors of order $O(\Delta t /a)^4$~\cite{Moore:2001vf}. For our parameter
choices, we found experimentally that this level of accuracy was crucial for the
real-time and Monte-Carlo parts of the code to agree sufficiently precisely on
the position of the separatrix.

\clearpage

\begin{widetext}
  
  \section*{Erratum}

The dynamical factor $A_\text{dyn}$ was missing in our one-loop results for the
nucleation rate. This was due to a typo in the script used to compute it. All our
lattice results are unaffected. The tree-level and LPA results are also unaffected.

Correctly including the dynamical factor in the one-loop rate brings it
in somewhat closer agreement with the lattice results. At the temperature simulated,
Eq.~\eqref{eq:rate_numbers_compared} becomes
\begin{equation}
  |\log\Gamma|^\text{lat} = 74.09(5),
  \quad
  |\log\Gamma|^\text{pt} = 38.0+29.1+\cdots .
\end{equation}
This correction reduces the discrepancy between lattice and one-loop
perturbation theory from 20\% to 9\%. Our broad conclusions, that perturbation
theory does significantly worse
in estimating the nucleation rate than the field condensate, are weakened
but not invalidated.
At the critical temperature, lattice and one-loop
perturbation theory agree on $\Delta \langle\phi\rangle_\text{c}$ to $0.2\%\ll 9\%$.
The different behaviour of perturbation theory for these two observables still
calls for an explanation.

Figure~\ref{fig:erratum} here is a corrected version of
Fig.~\ref{fig:rate} that includes the dynamical factor for the one-loop rate.
No other results are affected. The data have also been updated on Zenodo\footnote{\url{https://doi.org/10.5281/zenodo.10891523}}.

\begin{figure}[h]
  \includegraphics[width=0.4\textwidth]{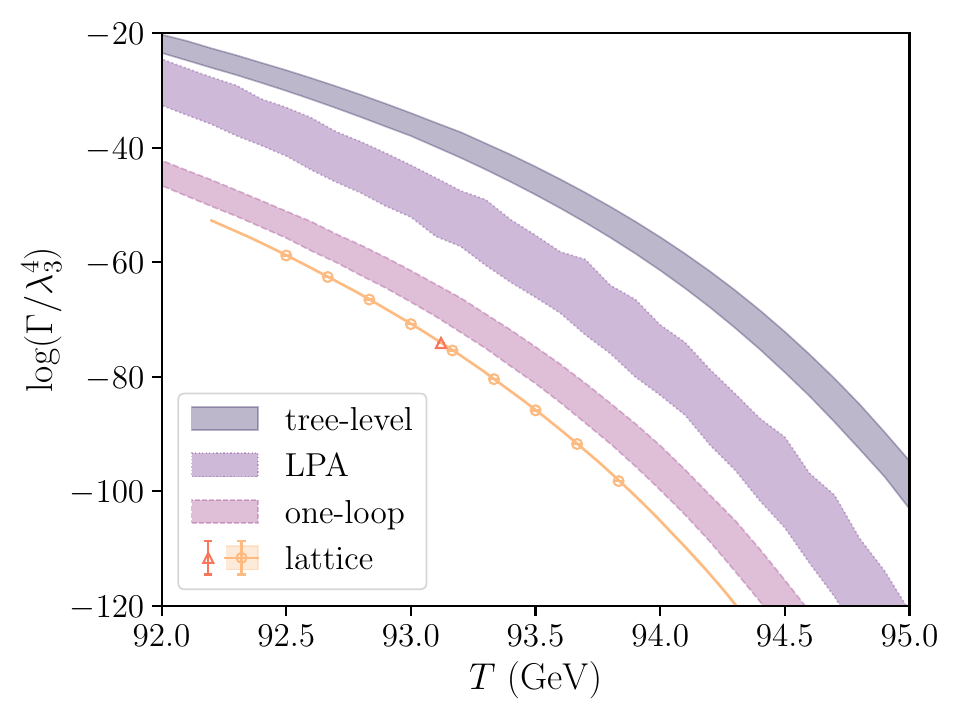}
  \caption{Corrected version of
  Fig.~\ref{fig:rate}. Note that only the one-loop results are changed, bringing
  them into somewhat closer agreement with the lattice results.\label{fig:erratum}}
\end{figure}

\end{widetext}

\end{document}